\documentclass[aip,graphicx, jmp, amsmath, amssymb, reprint]{revtex4-1}

\usepackage{natbib}
\usepackage{color}
\usepackage{graphicx}% Include figure files
\usepackage{dcolumn}% Align table columns on decimal point
\usepackage{bm}% bold math
\usepackage{hyperref}
\usepackage[all]{hypcap}
\hypersetup{colorlinks   = true}
\hypersetup{linkcolor=blue}
\hypersetup{citecolor=blue}
\hypersetup{urlcolor=black}

\draft % marks overfull lines with a black rule on the right

\begin{document}

\title{Phosphorus oxide gate dielectric for black phosphorus field effect transistors} %Title of paper

\author{W. Dickerson}
 \email{william.dickerson@mail.mcgill.ca}
\affiliation{Department of Electrical and Computer Engineering, McGill University, Montreal, Quebec, H3A 0E9, Canada}
\author{V. Tayari}
\affiliation{Department of Electrical and Computer Engineering, McGill University, Montreal, Quebec, H3A 0E9, Canada}
\author{I. Fakih}
\affiliation{Department of Electrical and Computer Engineering, McGill University, Montreal, Quebec, H3A 0E9, Canada}
\author{A. Korinek}
\affiliation{Department of Materials Science and Engineering, McMaster University, Hamilton, Ontario, L9H 4L7, Canada}
\author{M. Caporali}
\affiliation{Istituto Chimica dei Composti OrganoMetallici-CNR, Sesto Fiorentino, Italy}
\author{M. Serrano-Ruiz}
\affiliation{Istituto Chimica dei Composti OrganoMetallici-CNR, Sesto Fiorentino, Italy}
\author{M. Peruzzini}
\affiliation{Istituto Chimica dei Composti OrganoMetallici-CNR, Sesto Fiorentino, Italy}
\author{S. Heun}
\affiliation{NEST, Istituto Nanoscienze-CNR and Scuola Normale Superiore, Pisa, Italy}
\author{G. A. Botton}
\affiliation{Department of Materials Science and Engineering, McMaster University, Hamilton, Ontario, L9H 4L7, Canada}
\author{T. Szkopek}
 \email{thomas.szkopek@mcgill.ca}
\affiliation{Department of Electrical and Computer Engineering, McGill University, Montreal, Quebec, H3A 0E9, Canada}

\date{\today}

\begin{abstract}
The environmental stability of the layered semiconductor black phosphorus (bP) remains a challenge. Passivation of the bP surface with phosphorus oxide, PO$_x$, grown by a reactive ion etch with oxygen plasma is known to improve photoluminescence efficiency of exfoliated bP flakes. We apply phosphorus oxide passivation in the fabrication of bP field effect transistors using a gate stack consisting of a PO$_x$ layer grown by reactive ion etching followed by atomic layer deposition of Al$_2$O$_3$. We observe room temperature top-gate mobilities of 115 cm$^2$V$^{-1}$s$^{-1}$ in ambient conditions, which we attribute to the low defect density of the bP/PO$_x$ interface.
\end{abstract}


\maketitle %\maketitle must follow title, authors, abstract and \pacs

Black phosphorus (bP) is a direct band gap ($E_g = 0.3~\mathrm{eV}$) semiconductor with a puckered honeycomb layer structure characterized by van der Waals interlayer bonding \cite{morita1986semiconducting,keyes1953electrical}. The most thermodynamically stable allotrope of phosphorus, bP exhibits ambipolar conduction, anisotropic conductivity, and can be exfoliated down to the atomic monolayer limit \cite{li2014black,xia2014rediscovering,liu2014phosphorene,tayari2015two}. Exfoliation of bP in a nitrogen environment followed by encapsulation with hexagonal boron nitride in a vacuum environment has led to the observation of $\sim45,000~\mathrm{cm^2V^{-1}s^{-1}}$ hole mobility at cryogenic temperatures \cite{long}. Importantly, bP is subject to degradation by photo-oxidation with a reaction rate that increases as bP thickness decreases \cite{favron2015photooxidation}. A number of passivation techniques have been developed with varying degrees of success, including encapsulation with Al$_2$O$_3$ \cite{na2014few,wells2015passivation,kim2014toward,das2014ambipolar}, hexagonal boron nitride (h-BN) \cite{chen2015high,cao2015quality,avsar2015air,long}, polymer layers \cite{tayari2015two,passaglia} and functionalization with nickel nanoparticles \cite{caporali}. More recently, it has been demonstrated that the formation of a dense phosphorus oxide, PO$_x$, layer by oxygen plasma dry etching followed by Al$_2$O$_3$ deposition results in stable encapsulation of bP without compromising photoluminescence (PL) efficiency \cite{pei2016producing}. The preservation of PL efficiency indicates that the interface between bP and PO$_x$ does not measurably increase non-radiative recombination rates and is thus an effective surface passivation strategy.

In this work, we apply the PO$_x$ passivation approach of Pei \textit{et al.}\cite{pei2016producing} to fabricate top-gated bP field effect transistors (FETs). The use of a native oxide for passivation and gate stack formation in bP FETs is appealing as a direct analogue to the use of silicon oxide in silicon FET technology. Various phases of PO$_x$ are known, including a rhombohedral crystal of molecular P$_4$O$_{10}$ \cite{cruickshank1}, and the most thermodynamically stable form of P$_2$O$_5$ which is itself a layered material composed of a hexagonal network of edge connected $\mathrm{(PO)}_4^{3-}$ tetrahedra \cite{stachel}. The PO$_x$ layer passivates the bP surface, and acts as a seeding layer for subsequent atomic layer deposition of high-quality gate dielectrics such as Al$_2$O$_3$. We have fabricated dual-gate bP FETs, a bottom gate formed by a heavily doped, oxidized silicon substrate, and a top gate structure with a PO$_x$/Al$_2$O$_3$ dielectric stack. Room temperature top gate field effect mobilities of up to $115~\mathrm{cm^2V^{-1}s^{-1}}$ are achieved.

In our experiments, we used bP crystals prepared according to the procedure developed by Nilges \textit{et al.} \cite{nilges2008fast}, wherein high-purity red phosphorus ($>$ 99.99\%), tin ($>$ 99.999\%), and gold ($>$ 99.99\%) are heated in a muffle oven with a SnI$_4$ catalyst. The solid product was placed in a quartz tube, subjected to several evacuation-purge cycles with N$_2$ gas, and then sealed under vacuum. The evacuated quartz tube was heated to 406$^\circ$C at 4.2$^\circ$C/min, where it remained for 2 hours. The tube was then heated to 650$^\circ$C at 2.2$^\circ$C/min and held at this temperature for 3 days. The tube was then cooled at 0.1$^\circ$C/min. The final product is crystalline bP with a typical crystal size ranging from 0.3 mm $\times$ 3 mm to 3 mm $\times$ 10 mm. The bP crystals were mechanically exfoliated in a nitrogen glove box with H$_2$O and O$_2$ levels maintained at $<1~\mathrm{ppm}$ to minimize photo-oxidation during sample processing. The bP flakes were exfoliated onto either single crystal quartz substrates, or $5~\mathrm{m\Omega cm}$ As-doped, (001) Si substrates with 300~nm of dry thermal SiO$_2$. The SiO$_2$ surface was treated with a hexamethyldisilazane (HMDS) layer to suppress charge transfer doping \cite{lafkioti}.

\begin{figure}%[ht]
	\centering
	\includegraphics[width=.5\textwidth]{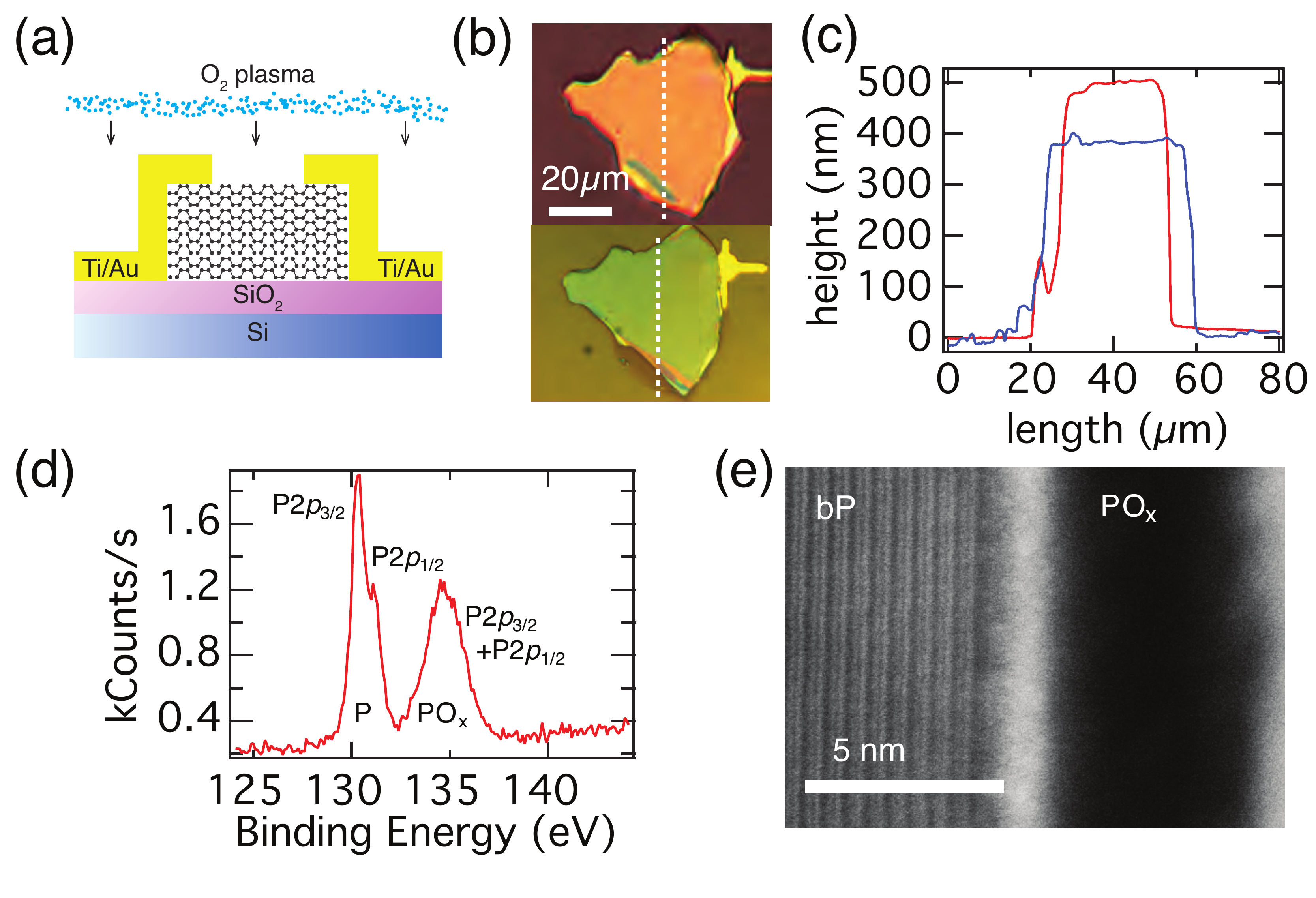}
	\caption{Oxidation of bP and characterization of the native oxide. (a) A schematic depiction of the oxidation process. (b) A freshly exfoliated bP flake before (top) and after (bottom) oxidation. White lines indicate profilometer traces taken before and after oxidation at approximately the same position to measure the etch rate. (c) Profilometer traces of the flake taken before (red) and after (blue) oxidation. The difference in bP width is an artifact arising from limited lateral alignment accuracy of the profilometer traces. (d) XPS P 2p core level spectrum of an oxidized bP surface indicating the presence of PO$_x$. (e) HAADF image of the top interface of an oxidized bP flake with PO$_x$ layer produced by RIE.}
	\label{fig:Oxidation}
\end{figure}

Amorphous PO$_x$ was grown on both bulk bP crystals and exfoliated bP flakes by reactive ion etching (RIE) in a custom built chamber with 10 sccm of O$_2$ flow at a chamber pressure of 200~mTorr with 300~W of RF bias power. Prior to oxidizing the bP samples, the RIE chamber was cleaned for 15~min with 10 sccm of O$_2$, 200~mTorr pressure and 300~W RF power. As discussed further below, the RIE process simultaneously oxidizes and etches the bP. We calibrated the RIE process by etching thick exfoliated flakes as shown in Fig. \ref{fig:Oxidation} a)-c). Stylus profilometer traces were taken before and after RIE processing, with the results of a 17~min etch shown in Fig. \ref{fig:Oxidation} b) and c). The estimated etch rate for our process is 0.98~\AA/s, which corresponds to an etch rate of 0.18~layers/s taking into account the 5.23~\AA~thickness per layer \cite{morita1986semiconducting}. By comparison, Pei \textit{et al.} report a comparable etch rate of 0.1~layers/s \cite{pei2016producing}. In our work, typical RIE processing times for thin ($< 1~\mathrm{\mu m}$) exfoliated bP flakes are 1-3 minutes, allowing for the growth of PO$_x$ with minimal etching.

To confirm the formation of PO$_x$, a freshly cleaved bP crystal was treated by RIE and analyzed by X-ray photoemission spectroscopy (XPS) with an Al K$_\alpha$ source. The resulting XPS spectrum of the P $2p$ core level is shown in Fig.\ref{fig:Oxidation}(d), showing contributions from both bP and PO$_x$. The P $2p_{3/2}$ peak at a binding energy of 130~eV and spin-orbit split P $2p_{1/2}$ peak are in good agreement with previous observations of bP XPS spectra\cite{pei2016producing,edmonds2015creating}. The overlapping P $2p_{1/2}$ and P $2p_{3/2}$ peaks at a binding energy of $\sim$134.5~eV are in remarkably good agreement with previous reports for P$_2$O$_5$ \cite{pei2016producing,edmonds2015creating,sherwood}. However, it remains a challenge to distinguish between the various phases of PO$_x$ \cite{gaskell}.

The oxidation process is best understood by the model described by Pei \textit{et al.} \cite{pei2016producing}. In the RIE process, O$_2$ plasma oxidizes the top layers of the bP, producing a dense PO$_x$ layer. Once this layer is established, subsequent etching by O$_2$ plasma penetrates the underlying bP, leading to further oxidation and PO$_x$ formation, while simultaneously sputtering away PO$_x$. A dynamic equilibrium is established between the sputtering of PO$_x$ and oxidation of the underlying bP, such that the PO$_x$ layer reaches a steady-state thickness. Pei \textit{et al.} \cite{pei2016producing} estimates a steady state thickness of approximately 10~nm after 40~s of oxidation, approximately twice the 5~nm thickness observed in our experiments (see below).

Notably, PO$_x$ is a strong hygroscopic desiccant and reacts violently with water. In order to suppress the reaction of PO$_x$ with water, we encapsulated the bP/PO$_x$ flakes with a 3~nm thick layer of Al$_2$O$_3$ by atomic layer deposition, exposing them to trimethylaluminum (TMA) for 148~seconds before proceeding with 25 ALD deposition cycles. By increasing exposure to hygroscopic TMA, the bP/PO$_x$ surface is in a dehydrated state before the introduction of H$_2$O into the ALD chamber to initiate oxidation of adsorbed TMA.

Transmission electron microscopy (TEM), was used to characterize the bP/PO$_x$/Al$_2$O$_3$ and bP/SiO$_2$ interfaces. An exfoliated bP sample was capped with a PO$_x$ layer grown for 3~min by RIE followed by a 3~nm layer of Al$_2$O$_3$ grown by ALD. A tungsten layer was deposited atop the flake for protection during the milling and polishing process. Electron energy loss spectroscopy (EELS) confirmed the presence of P, Al, and O. Fig. \ref{fig:Oxidation}(e) displays a high-angle annular dark-field (HAADF) image of the surface of the bP flake. The puckered honeycomb layers of P atoms in the bP crystal are clearly seen, as well as an amorphous layer of PO$_x$. We speculate that the PO$_x$ is an amorphous layer of molecular P$_4$O$_{10}$, but further work is required for unambiguous determination of the PO$_x$ phase. The PO$_x$ layer thickness is approximately $t\sim5~\mathrm{nm}$.

\begin{figure}%[ht]
	\centering
	\includegraphics[width=.5\textwidth]{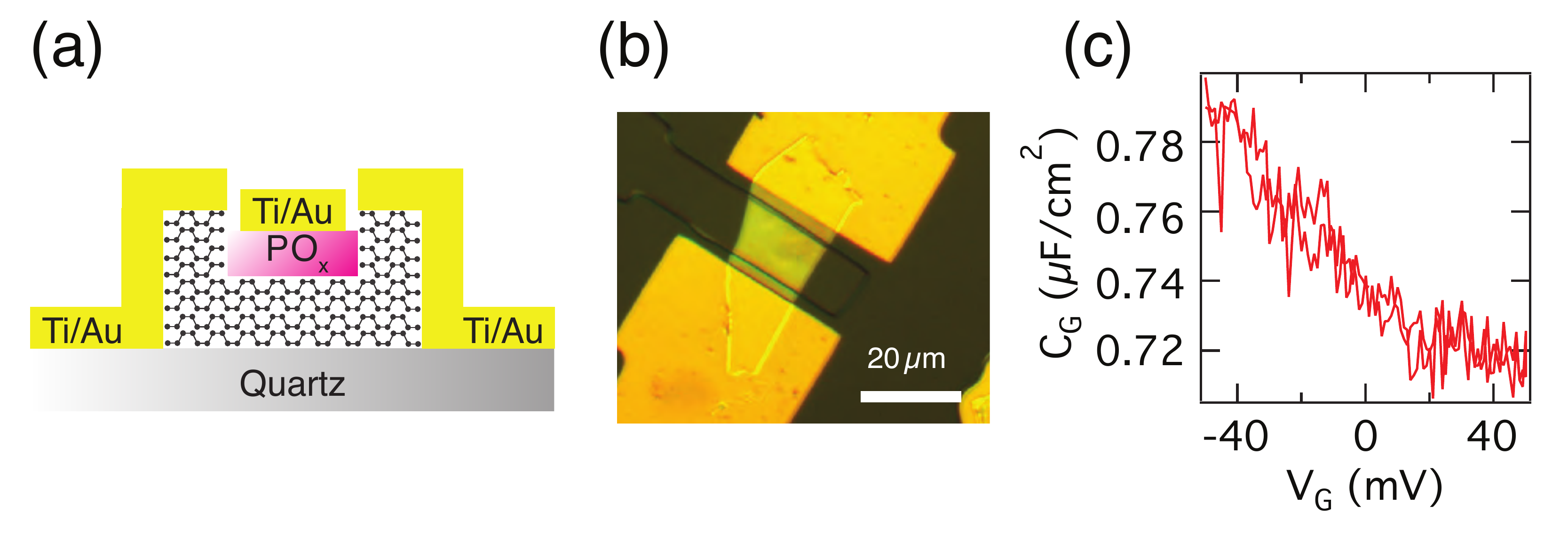}
	\caption{Capacitance measurement. (a) A schematic depiction of a bP FET on a quartz substrate used for measuring capacitance. (b) An optical micrograph taken before the final metalization step of the device used for capacitance measurement. The device used for the capacitance measurement had a $\sim$5 nm thick PO$_x$ capping layer grown by 3~minutes of RIE. (c) Capacitance versus top-gate voltage with an AC excitation of 1 mV peak-to-peak at a frequency of 1 MHz.}
	\label{fig:Quartz}
\end{figure} 

We measured the capacitance of the PO$_x$ layer alone as a gate dielectric with bP flakes fabricated on an insulating quartz substrate. A schematic of the bP device is displayed in Fig. \ref{fig:Quartz}(a). The bP flakes were first exfoliated on a quartz substrate and contacted with Ti/Au (5~nm / 80~nm) electrodes using an electron beam lithography process. The PO$_x$ was then deposited by 3~min of RIE. Finally, a second electron beam lithography process was used to deposit Ti/Au (5~nm / 80~nm) gate electrodes. An optical micrograph of a device before top-gate metallization is shown in Fig. \ref{fig:Quartz}(b). The top-gate capacitance per unit area $C_G$ versus top-gate dc bias voltage $V_G$ was measured with an AC excitation of $V_{ac}=1~\mathrm{mV}$ peak-to-peak at a frequency of 1.00~MHz, and is displayed in Fig. \ref{fig:Quartz}(c). At $V_G=0~\mathrm{V}$, the capacitance per unit area is $C_G=0.74~\mathrm{\mu F/cm^2}$. The gate capacitance is a series combination of: the capacitance $C_{PO_x}$ of the PO$_x$ layer; and the gate voltage dependent quantum capacitance $C_Q$ of the hole accumulation layer at the bP surface. The latter capacitance contribution is approximated from a non-degenerate thermodynamic density of states as $C_Q = e^2 \partial p / \partial E_F \sim e^2 p / k_B T = 6.2~\mathrm{\mu F/cm^2}$ for an estimated carrier density $p = 10^{12}\mathrm{/cm^2}$ at $T=300~\mathrm{K}$. Consequently, we find the PO$_x$ layer contribution $C_{PO_x} =( C^{-1}_G-C^{-1}_Q )^{-1}  = 0.84~\mathrm{\mu F/cm^2}$. For an estimated PO$_x$ layer thickness of $t=5$~nm determined from TEM images, the inferred dielectric constant of our PO$_x$ layer is $\epsilon = 19\epsilon_0$.

\begin{figure}%[ht]
	\centering
	\includegraphics[width=.5 \textwidth]{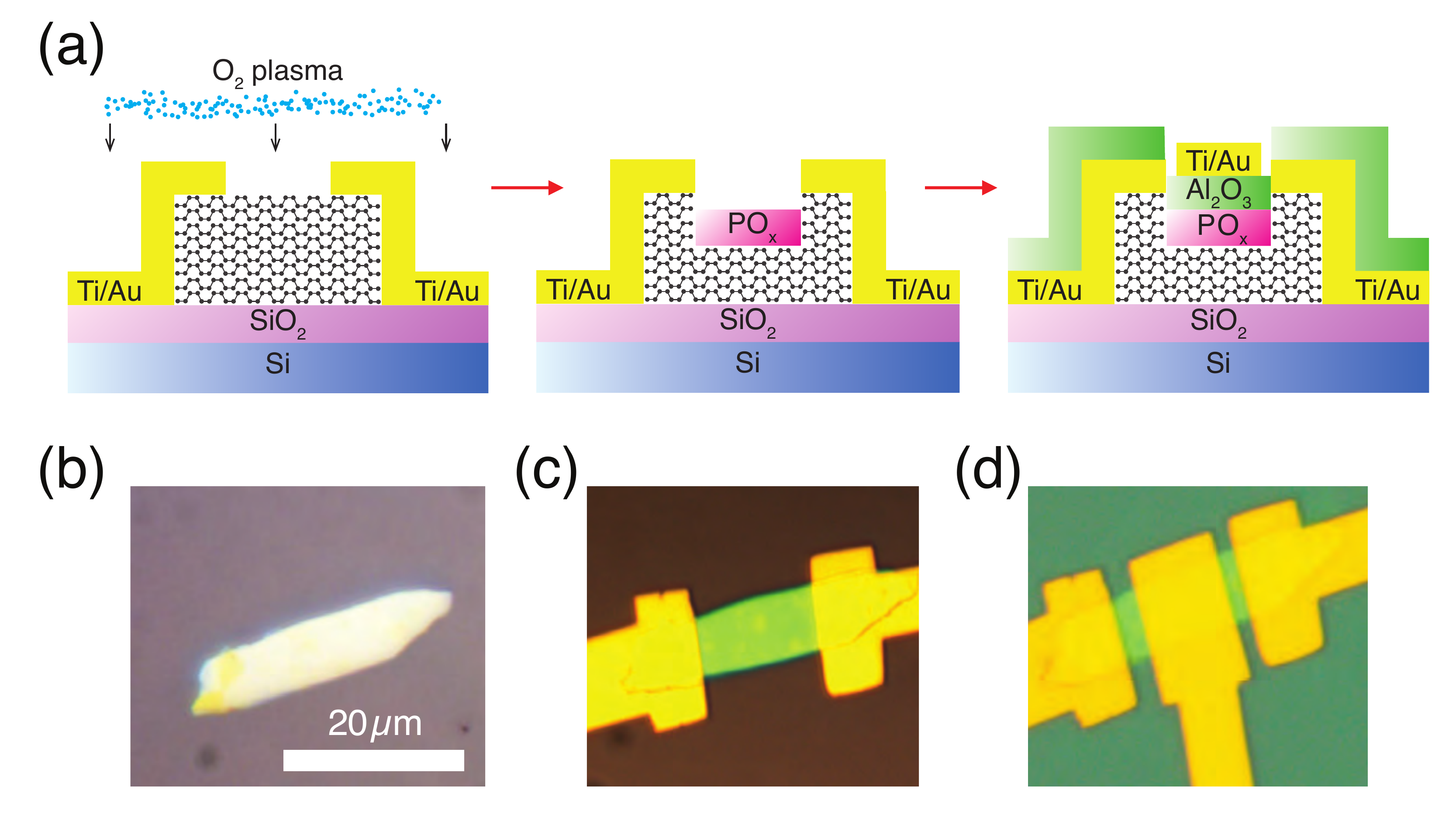}
	\caption{Dual-gate device preparation. (a) Schematic depiction of the oxidation and Al$_2$O$_3$ encapsulation process. (b) An optical micrograph of a freshly exfoliated bP flake. The scale bar is 20 $\mu$m. (c) Optical micrograph of the same flake after definition of the source and drain electrodes, 1 minute of RIE, and deposition of a 3~nm Al$_2$O$_3$ capping layer. (d) Top gate electrode definition.}
	\label{fig:FETs}
\end{figure}

We fabricated devices for field effect measurements in a dual gate configuration, similar to that first proposed by Sakaki for velocity modulated transistors \cite{sakaki1982velocity}, and recently used to compare top-gate and back-gate operation in bP FETs \cite{tayari2016dual}. A schematic of the fabrication process is shown in Fig. \ref{fig:FETs}(a). A freshly exfoliated bP flake on an SiO$_2$/Si substrate is displayed in Fig. \ref{fig:FETs}(b). A standard electron beam lithography process was used to define Ti/Au (5~nm / 80~nm) source-drain electrodes (Fig. \ref{fig:FETs}(c)). The PO$_x$ and Al$_2$O$_3$ layers were then deposited by 3~min of RIE and 25 cycles of ALD. Finally, a second electron beam lithography process was used to define Ti/Au (5~nm / 80~nm) top gate electrodes (Fig. \ref{fig:FETs}(d)). For further protection, bP FETs were encapsulated with polymethylmethacrylic (PMMA) layers. Taking into account the capacitance $C_{Al_2O_3}=\epsilon/t=9.34\epsilon_0/3~ \mathrm{nm} = 2.7~\mathrm{\mu F/cm^2}$ of the 3~nm layer of Al$_2$O$_3$, the top-gate capacitance is estimated to be $C_{TG} = ( C^{-1}_G+C^{-1}_{Al_2O_3} )^{-1} = 0.58~\mathrm{\mu F/cm^2}$. The back-gate capacitance is a significantly smaller $C_{BG} = \epsilon/t=3.9\epsilon_0/300~ \mathrm{nm}=11.5~\mathrm{n F/cm^2}$.

\begin{figure}%[ht]
	\centering
	\includegraphics[width=.45\textwidth]{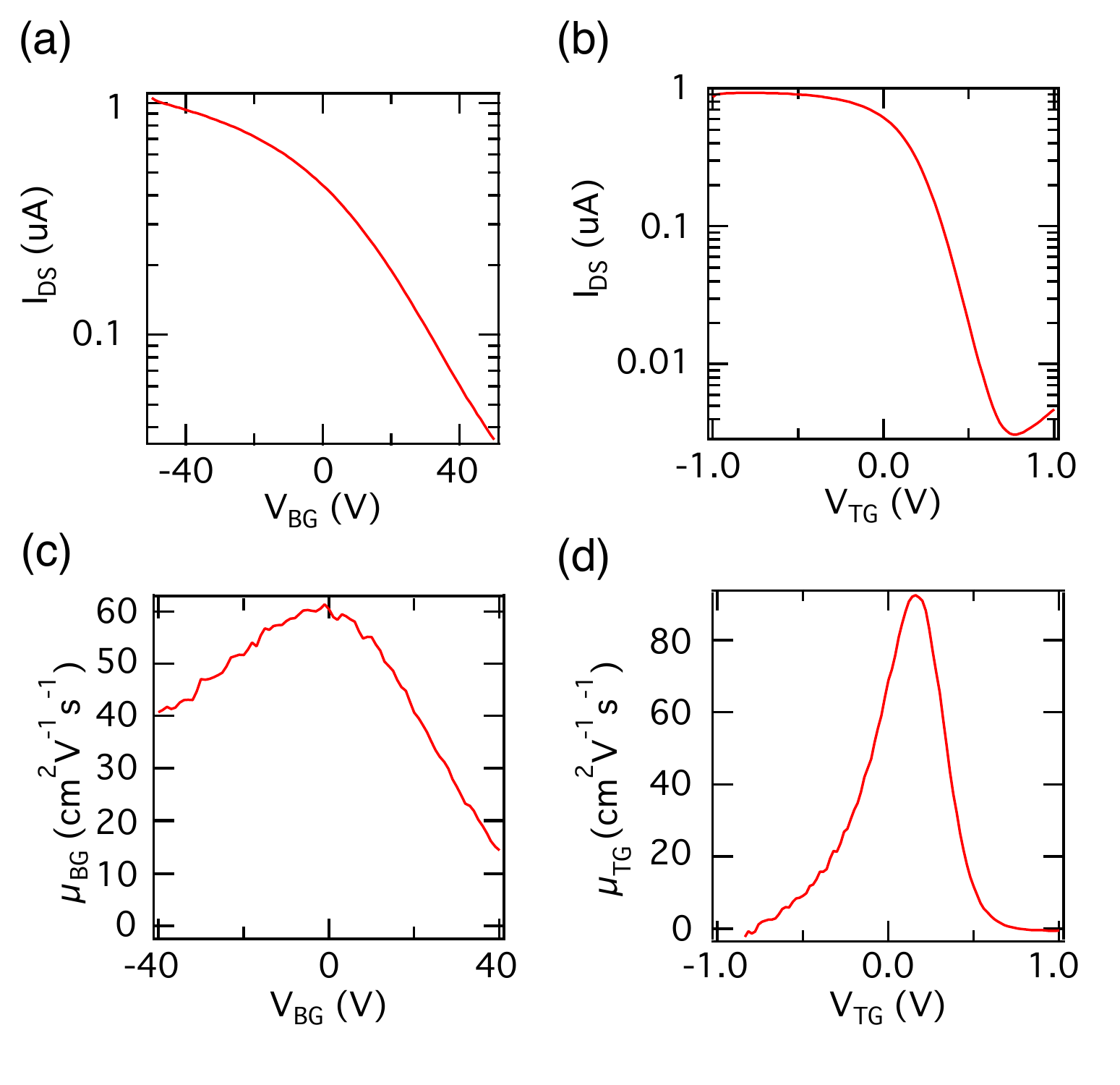}
	\caption{Dual-gate transistor characteristics. (a) Bottom gate sweep ($V_{DS}=50$ mV and  $V_{TG}=0$ V). (b) Top gate sweep ($V_{DS}=50$ mV and  $V_{BG}=0$ V). (c) Bottom gate mobility ($V_{DS}=50$ mV and  $V_{TG}=0$ V). (d) Top gate mobility ($V_{DS}=50$ mV and  $V_{BG}=0$ V).}
	\label{fig:Transport1}
\end{figure}

All electrical measurements were carried out in a vacuum probe station, $P<10^{-4}~\mathrm{Torr}$ using quasi-dc excitation with a semiconductor parameter analyzer. Leakage currents through the top and bottom oxide layers were measured through-out all measurements, and never exceeded 5\% of the measured source-drain current for all results shown here. Fig. \ref{fig:Transport1} a) and b) display the $I-V$ characteristics of our highest-mobility PO$_x$/Al$_2$O$_3$-passivated device. The source-drain current $I_{DS}$ versus bottom-gate voltage $V_{BG}$ with fixed top-gate voltage $V_{TG}=0~\mathrm{V}$ at a source-drain bias voltage $V_{DS}=50~\mathrm{mV}$ at a temperature of 300 K is plotted in Fig. \ref{fig:Transport1}(a). Similarly, $I_{DS}$ versus top-gate voltage $V_{TG}$ with fixed bottom-gate voltage $V_{BG}=0~\mathrm{V}$ at a source-drain bias voltage $V_{DS}=50~\mathrm{mV}$ is shown in Fig. \ref{fig:Transport1}(b). In both cases, hole conduction is observed, with an on/off current ratio for the bottom-gate and top-gate modulation of $I_{on}/I_{off}\sim30$ and $I_{on}/I_{off}\sim300$, respectively. Noting that the capacitance ratio $C_{TG}/C_{BG}\sim50$, the top-gate voltage range $-1~\mathrm{V} < V_{TG} < +1~\mathrm{V}$ and bottom-gate voltage range $-50~\mathrm{V} < V_{BG} < +50~\mathrm{V}$ extend over an induced charge displacement of approximately $\Delta D / e = C \Delta V / e = 7.2\times10^{12}\mathrm{cm^{-2}}$.

The field effect mobility is determined from the channel conductance $G_{DS} = I_{DS}/V_{DS}$ by $\mu_{X} = (L/W) ( \partial G_{DS} / \partial (C_{X} V_{X}) )$, where $L$ is the channel length, $W$ is the channel width and $X$ refers to either top-gate (TG) or bottom-gate (BG). The room temperature field effect mobilities are plotted in Fig. \ref{fig:Transport1}(c) and Fig. \ref{fig:Transport1}(d). The peak top-gate mobility of $\mu_{TG} = 90~\mathrm{cm^2V^{-1}s^{-1}}$ exceeds the peak back-gate mobility of $\mu_{BG} = 60~\mathrm{cm^2V^{-1}s^{-1}}$ in the same bP FET. The PO$_x$/Al$_2$O$_3$ top-gate is thus seen to give 10-fold improvement in on/off current ratio and a 50\% improvement in peak field effect mobility over the SiO$_2$ back-gate. In the case of both top-gate and back-gate modulation seen in Fig. \ref{fig:Transport1}(c) and (d), the field effect mobility falls to a negligible value at positive bias due to the depletion of hole density and subsequent loss of conductance modulation. At negative bias, channel conductance saturates due to the parasitic effects of access resistance and contact resistance, suppressing transconductance and leading to an under-estimation of field effect mobility. The saturation of conduction at negative bias is more prevalent with top-gate modulation than bottom-gate modulation because the top gate modulates a smaller area of the channel than the bottom gate, and thus the top-gate configuration suffers from higher access resistance. Despite the difference in access resistance, the field effect mobility extracted from raw conductance modulation data is superior in the top-gate configuration.

There are a number of physical mechanisms that may lead to an improvement in transfer characteristics in a top-gated hole accumulation layer: a reduced density of charged impurities at the top bP/PO$_x$ interface as compared to the bP/SiO$_2$ interface, stronger screening of charged impurities due to the shorter electrical length of the top-gate versus the bottom-gate, and a reduced contact and access resistance to the top accumulation layer where the source-drain electrodes are deposited as compared with the bottom accumulation layer. More subtle mechanisms, such as the role of phonon scattering at the respective bP/dielectric interfaces could also play a role.

\begin{figure}%[ht]
	\centering
	\includegraphics[width=.5\textwidth]{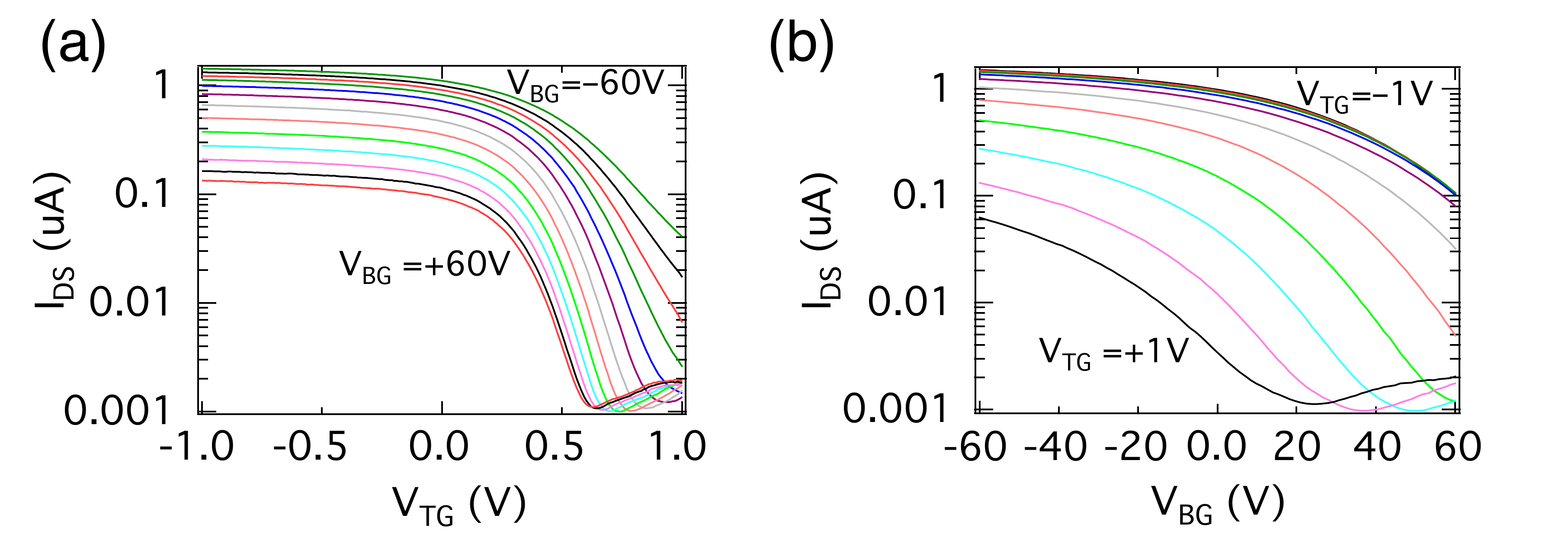}
	\caption{Dependence of bP FET conduction on both top- and bottom-gate voltages. (a) Top-gate sweeps taken at various bottom-gate biases. The bottom-gate bias was stepped in increments of 10 V. (b) Bottom-gate sweeps taken at various top-gate biases. The top-gate bias was stepped in increments of 0.2 V.}
	\label{fig:Transport2}
\end{figure} 

Lastly, we investigated bp FET channel modulation in a dual-gate mode, akin to the velocity modulated transistor proposed by Sakaki \cite{sakaki1982velocity}. The source-drain current $I_{DS}$ versus top-gate voltage $V_{TG}$ at a variety of bottom-gate voltages $V_{BG}$ is plotted in Fig. \ref{fig:Transport2}(a). Similarly, $I_{DS}$ versus bottom-gate voltage $V_{BG}$ at a variety of top-gate voltages $V_{TG}$ is plotted in Fig. \ref{fig:Transport2}(b). Several effects in the dual-gate mode are evident. Both the threshold voltage and on-current for hole accumulation induced by one gate voltage is modulated by the voltage applied to the other gate. The former effect is a consequence of both gate electrodes modulating the potential through-out the exfoliated bP flake. The latter effect is a consequence of gate modulation of both contact resistance and channel mobility, as has been previously observed in bP \cite{tayari2016dual}. Notably, the peak top-gate field effect mobility can be increased to $\mu_{TG}=115~\mathrm{cm^2V^{-1}s^{-1}}$ at a bottom-gate voltage $V_{BG}=-50~\mathrm{V}$.

The improvement in mobility of a hole accumulation layer at the bP/PO$_x$ interface in this work as compared to the bP/Al$_2$O$_3$ interface of previous work\cite{tayari2016dual} can be attributed to the combined action of etching away the bP surface damaged by oxidation under ambient conditions and the low defect density of the bP/PO$_x$ interface formed by RIE \cite{pei2016producing}. The RIE process etches away the bP surface exposed to ambient water and oxygen \cite{favron2015photooxidation}, allowing a clean bP/PO$_x$ interface to be formed. The PO$_x$ layer formed by RIE is itself an ideal seeding layer for subsequent ALD steps to enable gate dielectric engineering. In contrast, ALD of a top-gate dielectric such as Al$_2$O$_3$ directly atop bP is subject to the low quality of an ambient exposed bP surface.

In summary, we have shown that the PO$_x$/Al$_2$O$_3$ layers produced by RIE and ALD treatment of bP can be used as an effective gate dielectric stack in bP FETs with on/off current ratios $>300$ and field effect mobilities of $115~\mathrm{cm^2V^{-1}s^{-1}}$ at room temperature, exceeding the performance on the SiO$_2$ bottom-gate in the same bP FET channel. The simple PO$_x$/Al$_2$O$_3$ process enables the fabrication of dual-gated FETs, which have seen great interest in the context of the Stark effect, where band structure is modulated by applied electric field \cite{Deng2017, Yan2017}.

\textbf{Acknowledgements}
W.D., V.T., I.F. and T.S. acknowledge support from the Canadian Natural Sciences and Engineering Research Council, Fonds Qu\'ebecois de Recherche - Nature et Technologies. The TEM work (HAADF imaging and EELS) was carried out at the Canadian Centre for Electron Microscopy, a facility supported by the Canada Foundation for Innovation under the MSI program, NSERC and McMaster University. M.C., M.S.-R., S.H. and M.P. express thanks to the European Research Council for funding the project PHOSFUN ÒPhosphorene functionalization: a new platform for advanced multifunctional materialsÓ (Grant Agreement No. 670173) through an ERC Advanced Grant to MP. S.H. acknowledges support from Scuola Normale Superiore, project SNS16 B HEUN-004155.

\end{document}